\documentclass[12pt]{article}
\usepackage{graphicx,amssymb,amsfonts,amsmath,epsfig,color}

\makeatletter
\DeclareSymbolFont{AMSb}{U}{msb}{m}{n}
\DeclareSymbolFontAlphabet{\mathbb}{AMSb}
\renewcommand{\section}{\@startsection{section}{1}{\z@}%
                                    {-7ex \@plus -1ex \@minus -.2ex}%
                                    {2.5ex \@plus.2ex}%
                                    {\normalfont\large\scshape\centering}}
\renewcommand{\subsection}{\@startsection{subsection}{2}{\z@}%
                                       {-5ex \@plus -1ex \@minus -.2ex}%
                                       {1.5ex \@plus.2ex}%
                                       {\normalfont\normalsize\scshape}}
\renewcommand{\subsubsection}{\@startsection{subsubsection}{3}{\z@}%
                                       {-5ex \@plus -1ex \@minus -.2ex}%
                                       {1.5ex \@plus.2ex}%
                                       {\normalfont\normalsize\scshape}}

\renewcommand\@seccntformat[1]{\ignorespaces\csname #1name\endcsname\space
                               \csname the#1\endcsname.\quad}   
\newdimen\captionmargin
\newdimen\captionindent
\newdimen\captionwidth
\newcommand{\captionfont}{\slshape}
\newcommand\@captionlabel[1]{\textsc{#1:}\space}
\long\def\@makecaption#1#2{%
  \vskip\abovecaptionskip
  \captionwidth\hsize
  \advance\captionwidth -2\captionmargin
  \sbox\@tempboxa{\@captionlabel{#1}\captionfont #2}%
  \ifdim \wd\@tempboxa >\captionwidth
    \ifdim\captionindent>\z@
      \advance\captionwidth -\captionindent
      \hskip\captionindent
    \fi
    \hskip\captionmargin
    \parbox[t]{\captionwidth}{\leavevmode\hskip-\captionindent
      \@captionlabel{#1}\captionfont #2}%
  \else
    \global \@minipagefalse
    \hb@xt@\hsize{\hfil\box\@tempboxa\hfil}%
  \fi
  \vskip\belowcaptionskip}
\def\eqnarray{%
   \stepcounter{equation}%
   \def\@currentlabel{\p@equation\theequation}%
   \global\@eqnswtrue
   \m@th
   \global\@eqcnt\z@
   \tabskip\@centering
   \let\\\@eqncr
   $$\everycr{}\halign to\displaywidth\bgroup
       \hskip\@centering$\displaystyle\tabskip\z@skip{##}$\@eqnsel
      &\global\@eqcnt\@ne$\;\hfil{##}$\hfil
      &\global\@eqcnt\tw@$\;\displaystyle{##}$\hfil\tabskip\@centering
      &\global\@eqcnt\thr@@ \hb@xt@\z@\bgroup\hss##\egroup
         \tabskip\z@skip
      \cr}
\ifcase \@ptsize
\or
\or
\fi
  \divide\@tempdima\baselineskip
  \@tempcnta=\@tempdima
\makeatother
\begin{document}

%
%

\renewcommand{\theequation}{\arabic{section}.\arabic{equation}}
\renewcommand{\thefigure}{\arabic{figure}}
\newcommand{\gapprox}{%
\mathrel{%
\setbox0=\hbox{$>$}\raise0.6ex\copy0\kern-\wd0\lower0.65ex\hbox{$\sim$}}}
\textwidth 165mm \textheight 220mm \topmargin 0pt \oddsidemargin 2mm
\def\ib{{\bar \imath}}
\def\jb{{\bar \jmath}}

\newcommand{\ft}[2]{{\textstyle\frac{#1}{#2}}}
\newcommand{\be}{\begin{equation}}
\newcommand{\ee}{\end{equation}}
\newcommand{\bea}{\begin{eqnarray}}
\newcommand{\eea}{\end{eqnarray}}
\newcommand{\Identity}{{1\!\rm l}}
\newcommand{\cx}{\overset{\circ}{x}_2}
\def\CN{$\mathcal{N}$}
\def\CH{$\mathcal{H}$}
\def\hg{\hat{g}}
\newcommand{\bref}[1]{(\ref{#1})}
\def\espai{\;\;\;\;\;\;}
\def\zespai{\;\;\;\;}
\def\avall{\vspace{0.5cm}}
\newtheorem{theorem}{Theorem}
\newtheorem{acknowledgement}{Acknowledgment}
\newtheorem{algorithm}{Algorithm}
\newtheorem{axiom}{Axiom}
\newtheorem{case}{Case}
\newtheorem{claim}{Claim}
\newtheorem{conclusion}{Conclusion}
\newtheorem{condition}{Condition}
\newtheorem{conjecture}{Conjecture}
\newtheorem{corollary}{Corollary}
\newtheorem{criterion}{Criterion}
\newtheorem{defi}{Definition}
\newtheorem{example}{Example}
\newtheorem{exercise}{Exercise}
\newtheorem{lemma}{Lemma}
\newtheorem{notation}{Notation}
\newtheorem{problem}{Problem}
\newtheorem{prop}{Proposition}
\newtheorem{rem}{{\it Remark}}
\newtheorem{solution}{Solution}
\newtheorem{summary}{Summary}
\numberwithin{equation}{section}
\newenvironment{pf}[1][Proof]{\noindent{\it {#1.}} }{\ \rule{0.5em}{0.5em}}
\newenvironment{ex}[1][Example]{\noindent{\it {#1.}}}

\thispagestyle{empty}


\begin{center}

{\LARGE\scshape Evaporation of (quantum) black holes and energy conservation
\par}
\vskip15mm

\textsc{R. Torres}
\par\bigskip
{\em
Department of Applied Physics, UPC, Barcelona, Spain.}\\[.1cm]
\vspace{5mm}

\textsc{F. Fayos}
\par\bigskip
{\em
Department of Applied Physics, UPC, Barcelona, Spain.}\\[.1cm]
\vspace{5mm}

\textsc{O. Lorente-Esp\'{i}n}
\par\bigskip
{\em
Department of Physics and Nuclear Engineering, UPC, Barcelona, Spain.}\\[.1cm]
\vspace{5mm}

\end{center}

\section*{Abstract}
We consider Hawking radiation as due to a tunneling process in a black hole were quantum corrections, derived from Quantum Einstein Gravity, are taken into account.
The consequent derivation, satisfying conservation laws, leads to a deviation from an exact thermal spectrum. 
The non-thermal radiation is shown to carry information out of the black hole. Under the appropriate approximation, a quantum corrected temperature is assigned to the black hole. The evolution of the quantum black hole as it evaporates is then described by taking into account the full implications of energy conservation as well as the back-scattered radiation. It is shown that, as a critical mass of the order of Planck's mass is reached, the evaporation process decelerates abruptly while the black hole mass decays towards this critical mass.

\vskip10mm
\noindent KEYWORDS: Black Holes, Hawking radiation, Information loss.

\vspace{3mm} \vfill{ \hrule width 5.cm \vskip 2.mm {\small
\noindent E-mail: ramon.torres-herrera@upc.edu, f.fayos@upc.edu, oscar.lorente-espin@upc.edu }}



\newpage
\setcounter{page}{1}



\setcounter{equation}{0}

\section{Introduction}
Based on results of quantum field theory on a fixed curved background (Schwarzschild's solution) Hawking showed in 1975 \cite{Haw75} that black holes radiate a thermal spectrum of particles and derived an exact expression for their entropy.
Only recently \cite{P&W} Hawking radiation has been derived taking into account the back-reaction effect of the radiation on the black hole thanks to the requirement of energy conservation. Moreover, the method proposed in \cite{P&W} corresponds with the heuristic picture most commonly proposed of pair creation near the horizon of the black hole and the corresponding tunneling of particles.

One of the most interesting features of the tunneling method is that it shows that new terms appear in the distribution function which deviate it from pure thermal emission, i.e. the standard Boltzmann distribution.
Since the claim of information loss in black holes \cite{Haw76} has as one of its pillars that black holes have an exact thermal spectrum, it seems that the deviation from thermality could have consequences for the \textit{information loss paradox}, i.e., the radiation could allow the information to escape the black hole.


Of course, this picture is incomplete since, in order to describe the last stages of black hole evaporation, one should take into account quantum gravity effects. A step in this direction was taken by Bonanno and Reuter  in \cite{B&RIS} by introducing an effective quantum spacetime for spherically symmetric black holes based on the Quantum Einstein Gravity approach. They did this by using the idea of the Wilsonian renormalization group \cite{Wilson} in order to study quantum effects in the Schwarzschild spacetime. Specifically, they obtained a \textit{renormalization group improvement} of the Schwarzschild metric based upon a scale dependent Newton constant $G$ obtained from the exact renormalization group equation for gravity \cite{Reuter} describing the scale dependence of the effective average action \cite{Wett}\cite{R&W}. Later, in \cite{B&RIV}, the same authors described the strict thermal evolution of the improved black hole by estimating Hawking's energy flux directly from Stefan-Boltzmann's law.

Our aim in this letter is to analyze the evaporation of a quantum black hole (specifically, the solution found in \cite{B&RIS}) thanks to the consideration of a tunneling process in its horizon and, consequently, satisfying energy conservation. This has to allow us to find the quantum corrections to the temperature of the quantum black hole under the appropriate approximations. On the other hand, our study of the evolution of the evaporating quantum black hole satisfying energy conservation will take into account the effect of the non-negligible backscattered radiation. This analysis is intended to shed some light into the escape of information from black holes throughout their complete evaporation process as well as into the study of the lasts stages of their evaporation.


The letter has been divided as follows. Section \ref{secISS} introduces the solution for the quantum black hole (the \textit{improved Schwarzschild spacetime}) and its main properties. In section \ref{sectun} we summarize black hole radiation according to the tunneling method in an extended improved Schwarzschild spacetime. In section \ref{secback} we consider the backscattering of the emitted radiation taking into account energy conservation. This allows us, in section \ref{seclum}, to evaluate the luminosity of a quantum black hole when energy conservation is imposed and to compare it with the standard `thermal' result. The evolution of an evaporating quantum black hole fulfilling energy conservation is treated in section \ref{secevap}. In section \ref{secescape} we analyze the escape of information throughout the evaporation process. Finally the results are discussed in section \ref{seccon}.

\section{Improved Schwarzschild solution}\label{secISS}

The \textit{renormalization group improved} Schwarzschild solution found by Bonanno and Reuter \cite{B&RIS} can be written as
\begin{equation}\label{RGISch}
ds^2=-\left(1-\frac{2 G(R) M}{R}\right) dt_S^2+\left(1-\frac{2 G(R) M}{R}\right)^{-1} dR^2+ R^2 d\Omega^2.
\end{equation}
where
\begin{equation}
G(R)=\frac{G_0 R^3}{R^3+\tilde{\omega} G_0 (R+\gamma G_0 M)}, \label{GR}
\end{equation}
$G_0$ is Newton's universal gravitational constant, $M$ is the mass measured by an observer at infinity and $\tilde{\omega}$ and $\gamma$ are constants coming from the non-perturbative renormalization group theory and from an appropriate ``cutoff identification", respectively.
Despite the preferred value for $\gamma$ is $\gamma=9/2$, it is argued \cite{B&RIS}\cite{B&RIV} that the qualitative properties of this solution are fairly insensitive to the precise value of this constant.
In fact, the important differences appear only near the singularity. For instance, for the value $\gamma=9/2$ the usual singularity in the classical Schwarzschild solution does not exist in the improved solution while if, in order to simplify the calculations, one chooses $\gamma=0$ there is still a scalar curvature singularity at $R=0$, even if it has a milder character than in the classical case \cite{impcoll}. On the other hand, $\tilde \omega$ can be found by comparison with the the standard perturbative quantization of Einstein's gravity (see \cite{Dono} and references therein). It can be deduced that its precise value is $\tilde \omega=167/30\pi$, but again the properties of the solution do not rely on its precise value as long as it is strictly positive. A relevant fact with regard to $\tilde \omega$ is that it carries the quantum modifications. In effect, if we make explicit Planck's constant in (\ref{GR}), it can be considered that $\tilde \omega =167 \hbar/30\pi $ and, thus, $\tilde \omega=0$ would turn off the quantum corrections.

The horizons in this solution can be found by solving
\[
1-\frac{2 G(R) M}{R}=0.
\]
The number of positive real solutions to this equation correspond to the positive real solutions of a cubic equation and depends on the sign of its discriminant or, equivalently, on whether the mass is bigger, equal or smaller than a critical value $M_{cr}$. In general, the critical value takes the form
\begin{equation}
M_{cr}=a(\gamma) \sqrt{\frac{\tilde{\omega}}{G_0}}=a(\gamma) \sqrt{\tilde{\omega}} m_p \sim \sqrt{\tilde{\omega}} m_p,
\end{equation}
where $m_p$ is the Planck's mass and the function $a(\gamma)$ has, in general, an involved expression that, for reasonable values of $\gamma$ satisfies $a(\gamma)\sim 1$. In particular, the preferred value $\gamma=9/2$ provide us with
\[
M_{cr}=\frac{1}{24} \sqrt{\frac{1}{2} (2819+85 \sqrt{1105})}\sqrt{\frac{\tilde\omega}{G_0}}\simeq 2.21 \sqrt{\tilde{\omega}} m_p \simeq 2.94 m_p,
\]
while the value $\gamma=0$ implies
\[
M_{cr}=\sqrt{\frac{\tilde\omega}{G_0}}\simeq 1.33 m_p.
\]

If $M>M_{cr}$ then the equation has two positive real solutions $\{R_-,R_+\}$ satisfying $R_-<R_+$.
The inner solution $R_-$ represents a novelty with regard to the classical solution, while the outer solution $R_+$ can be considered as the \textit{improved Schwarzschild horizon}, i.e., the Schwarzschild horizon when the quantum modifications are taken into account. The `improvement' in this horizon can be made apparent for masses much bigger than Planck's mass if one expands $R_+$ in terms of $m_p/M$ obtaining
\[
R_+\simeq 2 G_0 M \left[1-\frac{(2+\gamma)}{8} \tilde\omega\ \left(\frac{m_p}{M}\right)^2\right].
\]
On the other hand, if $M=M_{cr}$ then there is only one positive real solution to the cubic equation, whereas if $M<M_{cr}$ the equation has not positive real solutions.

\section{Tunneling}\label{sectun}

Let us now consider Hawking radiation coming out from an improved black hole satisfying $M>M_{cr}$ thanks to the tunneling process occurring through the outer horizon $R_+$. First, we will rewrite the improved Schwarzschild's solution in Painlev\'e-like coordinates \cite{Pain} so as to have coordinates which are not singular at the horizon. In order to do this it suffices to introduce a new coordinate $t$ replacing the Schwarzschild-like time $t_S$ such that $t=t_S+h(R)$ and fix $h(R)$ by demanding the constant time slices to be flat. In this way one gets:
\begin{equation}
ds^2=-\left(1-\frac{2 G(R) M}{R}\right) dt^2+2 \sqrt{\frac{2 G(R) M}{R}} dt dR+ dR^2 + R^2 d\Omega^2,
\end{equation}
where $R$ can now take the values $0<R<\infty$.
In these coordinates the radial null geodesics describing the evolution of \emph{test} massless particles are given by
\begin{equation}\label{geodtest}
\frac{dR}{dt}=\pm 1-\sqrt{\frac{2 G(R) M}{R}}
\end{equation}
with the upper (lower) sign corresponding to outgoing (ingoing, respectively) geodesics.
Since the coefficients of the metric do not depend on $t$ there is a killing vector $\partial/\partial t$ which is straightforwardly found to be timelike for $R>R_+$, lightlike for $R=R_+$ (the event horizon) and spacelike for $R_-<R<R_+$\footnote{For the sake of completeness, let us comment that the killing vector is also lightlike for $R=R_-$ and timelike for $R<R_-$.}.
The possibility of tunneling is based on the fact that the killing vector is spacelike beneath the event horizon, what allows the existence of negative energy states. Pair production can occur either just inside the horizon with a positive energy particle tunneling out or
just outside the event horizon with a negative energy particle tunneling in.


In \cite{K&W}\cite{P&W} it was found that, when a self-gravitating shell of energy $E$ travels in
a spacetime characterized by an ADM mass $M$,
the geometry outside the shell is also characterized by $M$, but energy conservation implies that
the geometry inside the shell is characterized by $M-E$. It was also found that the shell
then moves on the geodesics given by the interior line element.
In this way, according to (\ref{geodtest}), one expects a shell of energy $E$ to satisfy the evolution equation
\begin{equation}\label{geodshell}
\frac{dR}{dt}=\pm 1-\sqrt{\frac{2 G(R;E) (M-E)}{R}}
\end{equation}
where $G(R;E)$ is the function $G(R)$ with $M$ replaced by $M-E$\;\; \footnote{Note that, from now on, the nomenclature `$;E$' means `$M$ should be replaced by $M-E$' is used for all the functions appearing in this letter whose dependence on $E$ is explicited.}.

%

Let us first consider pair production occurring just beneath the event horizon with the positive energy particle tunneling out.
The standard results of the WKB method for the tunneling through a potential barrier that would be classically forbidden can be directly applied due to the infinite redshift near the horizon \cite{P&W}. In particular, the semiclassical emission rate will be given by $\Gamma \sim \exp\{-2 \mbox{Im} S\}$, where $S$ is the particle action. Therefore, we have to compute the imaginary part of the action for an
outgoing positive energy particle which crosses the horizon $R_+$ outwards from $R_{in}$ to $R_{out}$.
\begin{equation}
\mbox{Im} S=\mbox{Im} \int_{R_{in}}^{R_{out}} p_R dR= \mbox{Im}  \int_{R_{in}}^{R_{out}}  \int_{0}^{p_R} dp'_R dR.
\end{equation}
Using Hamilton's equation $\dot{R}=+dH/dp_R\rfloor_R$ and $H=M-E'$, this can be written with the help of (\ref{geodshell}) as
\begin{eqnarray}\label{imS}
\mbox{Im} S&=& \mbox{Im} \int_{M}^{M-E} \int_{R_{in}}^{R_{out}} \frac{dR}{\dot R}  dH=\nonumber\\
&=&\mbox{Im} \int_{0}^{E} \int_{R_{in}}^{R_{out}} \frac{dR}{1-\sqrt{\frac{2 G(R;E') (M-E')}{R}}} (-dE').
\end{eqnarray}
If we define the functions $f(R;E')$ and $g(R;E')$ such that
\begin{displaymath}
f(R;E')\equiv 1-\frac{2 G(R;E') (M-E')}{R}=(R-R_+(E')) g(R;E'),
\end{displaymath}
where $R_+(E')$ is the position of the outer horizon when $M$ is replaced by $M-E'$ and $g$ satisfies
\[
g(R_+;E')=
\frac{\partial f(R;E')}{\partial R}\rfloor_{R=R_+(E')}\neq 0,
\]
then
by deforming the contour of integration so as to ensure that positive energy solutions decay in time and taking into account that a particle just inside the horizon tunnels just outside a shrunken horizon ($R_{in}>R_{out}$) one gets
\[
 \int_{R_{in}}^{R_{out}} \frac{dR}{1-\sqrt{\frac{2 G(R;E') (M-E')}{R}}} =-i \pi \frac{2}{g(R_+;E')}.
\]

We can then write (\ref{imS}) as
\begin{equation}\label{partchan}
\mbox{Im} S= \int_{0}^{E}  \frac{2 \pi}{g(R_+;E')} dE'\ .
\end{equation}

Tunneling also happens when a pair is created outside the horizon and the negative energy particle tunnels into the black hole.
Then, following the procedure for the Schwarzschild case in \cite{P&W}, the imaginary part of the action for this ingoing particle satisfies
\begin{equation}
\mbox{Im} \int_{0}^{-E} \int_{R_{out}}^{R_{in}} \frac{dR}{-1+\sqrt{\frac{2 G(R;M\rightarrow M+E')(M+E')}{R}}} dE'=\int_{0}^{E}  \frac{2 \pi}{g(R_+;E')} dE',
\end{equation}
what coincides with the result for the `particle channel' (\ref{partchan}).
Both particle and antiparticle tunneling contribute to the rate of the Hawking process, but we have seen that both contributions provide us with the same exponential term for the semiclassical rate
\begin{equation}\label{emprob}
\Gamma\sim e^{-2 \mbox{\scriptsize Im} S }=\exp\left(-4\pi \int_{0}^{E}  \frac{dE'}{g(R_+;E')} \right).
\end{equation}

When quadratic terms are neglected
we can develop Im $S$ up to first order in $E$ as
\[
\mbox{Im} S \simeq -\frac{2 \pi}{g(R_+,0)} E
\]
obtaining a thermal radiation for the quantum black hole ($\Gamma \sim \exp\{-E/T_{QBH}\}$) with temperature
\begin{equation}\label{TQBH}
T_{QBH}=\frac{g(R_+,0)}{4 \pi}=\frac{1}{4 \pi}\left. \frac{\partial f}{\partial R}\right\rfloor_{R=R_+}.
\end{equation}
This coincides with the expected temperature obtained by computing it as $T=\kappa/(2 \pi)$, where $\kappa=f'(R_+)/2$ is the standard surface gravity of the black hole of mass $M$.

Even if the explicit form of $g(R_+;0)$ with $\gamma\neq 0$ is rather complex, one can get an idea of the quantum gravity modifications to the tunneling for the Schwarzschild case \cite{P&W} ($\tilde\omega=0$) when the black hole mass is much bigger than Planck's mass
by performing an expansion in terms of Planck's constant or, in other words, of $\tilde\omega$, obtaining

%

\begin{equation}\label{TempAprox}
T_{QBH}\simeq \frac{1}{8 \pi G_0 M}-\frac{(1+\gamma) \tilde{\omega}}{32 \pi G_0^2 M^3},
\end{equation}
where the first term is the standard Hawking-Bekenstein temperature and the second term corresponds to the quantum modification to the temperature associated with the improvements to Schwarzschild's solution. These quantum modifications are more relevant if the black hole mass is of the order of the critical mass and, in fact, the temperature even becomes zero if the mass equals the critical mass. Notwithstanding these results about the temperature of the black hole, it is important to remark that the higher order terms in $E$, neglected in (\ref{TQBH}) and (\ref{TempAprox}), imply a deviation from pure thermal emission.

\section{Backscattering and energy conservation}\label{secback}

Whenever a wave is radiated from the black hole horizon its wave function satisfies a wave equation with an effective potential that depends on $R$ and the wave's angular momentum. This potential represents a barrier to the outgoing radiation, so that part of the radiation does not reach the future null infinity, but it is back-scattered. One can then  define the usual reflection $r_l(E)$ and transmissions $t_l(E)$ coefficients \cite{deWitt} for the scattering. In this way, it can be shown that the distribution function $<n(E, l)>$ for the Hawking radiation will be modulated as seen from infinity by a factor $\beta_{E  l}\equiv|t_l(E)|^2$ that is called the grey-body factor \cite{FN-S}. Moreover, without taking into account energy conservation, it can be shown
(see, for example, \cite{HNS}) that for any static spherically symmetric black hole with outer horizon $R_+$, and whenever $E M\ll 1$, the main contribution to the grey-body factor comes from the zero angular momentum $l=0$ and takes the form
\begin{equation}\label{gammast}
\beta\equiv\beta_{E 0}=4 E^2 R_+^2.
\end{equation}
For black hole masses much bigger than Planck's mass we have $R_+\simeq 2 G_0 M$ so that the grey-body factor takes the usual form \cite{St&Ch} for a Schwarzschild black hole
\begin{equation}\label{gammasch}
\beta_{Schw.}=16 G_0^2 E^2 M^2.
\end{equation}

The `standard' approximated grey-body factor neglects the effect of energy conservation and its subsequent back-reaction on the metric. However, according to our philosophy in this letter we would be interested in finding a better approximation that does take energy conservation into account.
As we stated in the previous section, in \cite{K&W}\cite{P&W} it was shown that whenever an outgoing self-gravitating shell of energy $E$ travels in a spacetime of total ADM mass $M$
energy conservation implies that the shell moves on the null geodesics given by the interior line element with mass $M-E$. On the other hand, this result can be extrapolated to ingoing shells (since a spacetime with an ingoing shell can be interpreted as the time-reversal of a spacetime with an outgoing shell). In this way, if an outgoing shell is back-scattered, thus becoming ingoing, the shell will have always moved on null geodesics given by a line element with mass $M-E$. Therefore, in order to obtain the modified approximated grey-body factor it suffices to repeat the usual backscattering calculations, but taking as the black hole mass $M-E$ instead of the standard choice $M$. With this, one just gets the straightforward result --in view of (\ref{gammast})--:
\begin{equation}\label{gammaEC}
\beta_{EC}=4 E^2 R_+(E)^2,
\end{equation}
where, as previously, $R_+(E)$ is $R_+$ with $M$ replaced by $M-E$.

\section{Spectral distribution and luminosity}\label{seclum}
Let us recall that neglecting quantum corrections and energy conservation we would have the standard thermal distribution for photons
\begin{equation}\label{nStand}
<n(E)>_{Stand.}=\frac{1}{\exp(8 \pi G_0 E M)-1}
\end{equation}
and the total flux, including back-scattering \cite{FN-S},
\begin{eqnarray}
L_{Stand.}\simeq \frac{1}{2 \pi}\int_0^\infty <n(E)>_{Stand.}  \beta_{Schw.}(M,E) E dE \nonumber\\
=\frac{1}{2 \pi}\int_0^\infty \frac{16 G_0^2 M^2 E^3 }{\exp(8 \pi G_0 M E)-1} dE = \frac{1}{7680 \pi G_0^2 M^2}.\label{LStand}
\end{eqnarray}
It is important to remark that the effect of backscattering cannot be neglected since the calculation without taking into account this factor result in total luminosities for the black hole around ten times bigger. Note also that, the usual approximation here \cite{FN-S} is to keep the form of the greybody factor in all the range of $E$ (even if, strictly speaking, it should satisfy $\beta_{Schw.}\leq 1$) since the distribution function decreases exponentially as $E$ grows.

On the other hand, if we consider the full consequences of energy conservation, the distribution function for the emission of photons
can be written as (see \cite{K&K} --correcting the result in \cite{K&W}--)
\begin{displaymath}
<n(E)>=\frac{1}{\exp \left(2 \mbox{Im} S \right)-1}.
\end{displaymath}
What for our quantum corrected solution becomes
\begin{equation}\label{nE}
<n(E)>=\frac{1}{\exp \left(4\pi \int_{0}^{E}  \frac{dE'}{g(R_+;E')} \right)
-1}.
\end{equation}
In order to compare this with the standard result (\ref{nStand}) note that if, and only if, the black hole mass is much bigger than Planck's mass, we can write it approximately as
\begin{displaymath}
<n(E)>\simeq\frac{1}{\exp \left\{8 \pi G_0 E \left(M-\frac{E}{2}\right)-2 \pi (1+\gamma) \tilde\omega \ln\left(1-\frac{E}{M}\right)\right\}-1}.
\end{displaymath}

According to our comments in section \ref{secISS}, there are no horizons if the black hole mass is smaller than $M_{cr}$. Therefore, $f(R;E')=0$ has no positive real solutions if the black hole mass (here $M-E'$) is smaller than  $M_{cr}$. In other words, $g(R;E')$ would be real if, and only if, the energy of the emitted particle $E'$ satisfies $E'\leq M-M_{cr}$.
This is very interesting since it imposes energy conservation by forbidding the emitted quantum to carry more energy than the black hole mass. Moreover, using (\ref{nE}), the luminosity modulated according to the greybody factor $\beta_{EC}$ has to be written as
\begin{eqnarray}
L(M)\simeq \frac{1}{2 \pi}\int_0^{M-M_{cr}} <n(E)>  \beta_{EC}(M,E) E dE \nonumber\\
=\frac{1}{2 \pi}\int_0^{M-M_{cr}} \frac{4 E^3 R_+(E)^2}{\exp\left(4\pi \int_{0}^{E}  \frac{dE'}{g(R_+;E')} \right)-1} dE,\label{lumi}
\end{eqnarray}
where we are taking into account in the integration limits that
the maximum energy of a radiated particle could be
$M-M_{cr}$. (In fact, this can be taken as an indication that a thermal spectrum, which would contain a tail of arbitrarily high energies, can not provide us with the correct spectrum).

\begin{figure}
\includegraphics[scale=1]{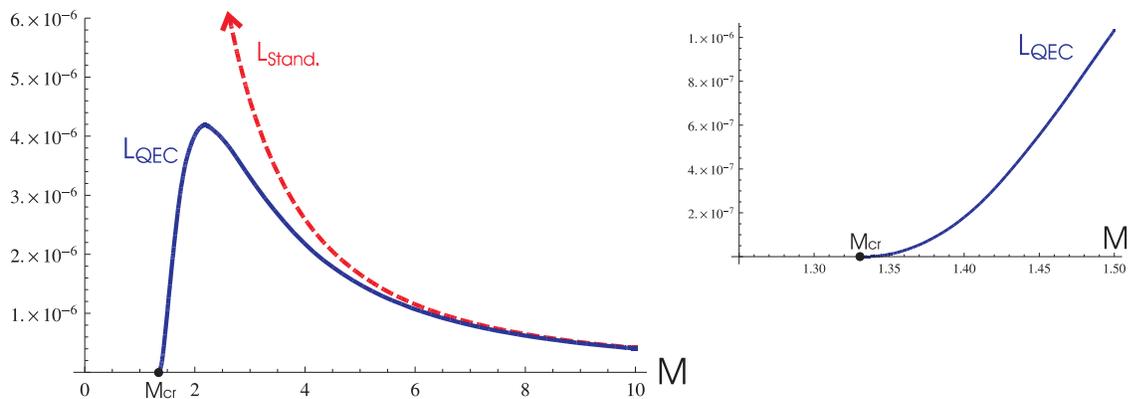}
\caption{\label{luminos} To the left the luminosity of a black hole as a function of its mass (taking into account backscattering). Quantum corrections (with $\gamma=0$) and energy conservation have been considered in order to draw the solid line $L_{QEC}$, while the dashed line $L_{Stand.}$ has been drawn considering the standard Schwarzschild background and neglecting energy conservation. As can be seen, in the standard case, if energy conservation is neglected the luminosity diverges as the black hole approaches its total evaporation. On the other hand, quantum corrections and energy conservation imply that the luminosity reaches a maximum for a non-null mass while it is zero for $M<M_{cr}$. A detail of the quantum corrected luminosity around $M_{cr}$ is shown to the right.}
\end{figure}

It is important to note that for masses much bigger than Planck's mass the integrand in (\ref{lumi}) is very similar to the integrand that appears without considering quantum corrections and energy conservation (\ref{LStand}).
%
%
Therefore, with regard to the total luminosity as seen from infinity, one expects the results of its computation to be very similar for macroscopic black holes regardless of whether quantum corrections and energy conservation are taken into account (\ref{lumi}) or not (\ref{LStand}). For these macroscopic black holes, the luminosity should grow as the mass of the black hole decreases. The differences become important as the lasts stages of the evaporation process are reached (see figure \ref{luminos}), since the luminosity in the standard `thermal' approach keeps increasing while, considering quantum corrections and energy conservation, the luminosity reaches a maximum and then decreases as the black hole's mass keeps decreasing (whenever its mass is bigger than the critical mass). Furthermore, the luminosity for an improved black hole with mass equal to the critical mass would be zero.
%
%
%
%
This sharply contrasts with the standard result (\ref{LStand}) that provides bigger luminosities for smaller masses and even implies that the luminosity diverges as the black hole's mass tends to zero, what should be considered as a nonsensical result that, even if quantum corrections were not considered, would be avoided by not allowing the black hole to emit particles with energies bigger than its own mass.

\section{Evaporating model}\label{secevap}

In order to modelize the evaporating black hole, let us first write the improved Schwarzschild's metric (\ref{RGISch}) in terms of ingoing Eddington-Finkelstein-like coordinates $\{u,R,\theta,\varphi\}$, where
\[
u=t_S+\int^R \frac{dR'}{1-2 G(R') M/R'}\ ,
\]
as
\begin{equation}\label{ScEF}
ds^2=-\left(1-\frac{2 G(R) M}{R}\right) du^2+2 du dR + R^2 d\Omega^2.
\end{equation}
This solution does not reflect the back-reaction associated to the lost of mass due to the tunneling effect. However, we can modelize the mass lost taking into account that, whenever a pair of virtual particles is created, when the particle with positive energy escapes to infinity its companion, with negative energy, falls into the black hole thus reducing its mass. In this way, if we consider negative energy massless particles following ingoing null geodesics $u=$constant, the mass of the black hole becomes a decreasing function $M(u)$. The metric which incorporates the effect of the decreasing BH mass due to the ingoing null radiation
is (\ref{ScEF}) with $M$ replaced by $M(u)$, i.e., it corresponds to an \textit{improved} ingoing Vaidya solution \cite{B&RIV}
\begin{equation}\label{Vaid}
ds^2=-\left(1-\frac{2 G(R; M(u)) M(u)}{R}\right) du^2+2 du dR + R^2 d\Omega^2.
\end{equation}
On the other hand, the flux of negative energy particles directed towards the black hole equals the flux of outgoing radiated particles and, therefore,
\begin{equation}\label{difM}
\frac{d M(u)}{du}=- L(M(u)).
\end{equation}
If we add to this picture that part of the (positive energy) outgoing radiation is backscattered (a fact that is reflected in the greybody factor $\beta$) we should use in this equation $L(M(u))$ modulated by the backscattered radiation.
For instance, one can consider the evolution corresponding to an evaporating BH taking into account the backscattered radiation, but neglecting quantum gravity corrections and the consequences of energy conservation, if we consider the luminosity as given in (\ref{LStand}). The mass evolution is then found with (\ref{difM}) by solving
\begin{displaymath}
\frac{d M(u)}{du}=-\frac{1}{7680 \pi G_0^2 M(u)^2}.
\end{displaymath}
If one sets the initial mass at $u=0$ such that $M(u=0)\equiv M_0$ then the mass evolution follows
\begin{equation}
M(u)=\sqrt[3]{M_0^3-u/(2560 \pi G_0^2)}.\label{MassScat}
\end{equation}
So that
the total evaporation is reached at  $u=2560 \pi G_0^2 M_0^3$.

\begin{figure}
\includegraphics[scale=1]{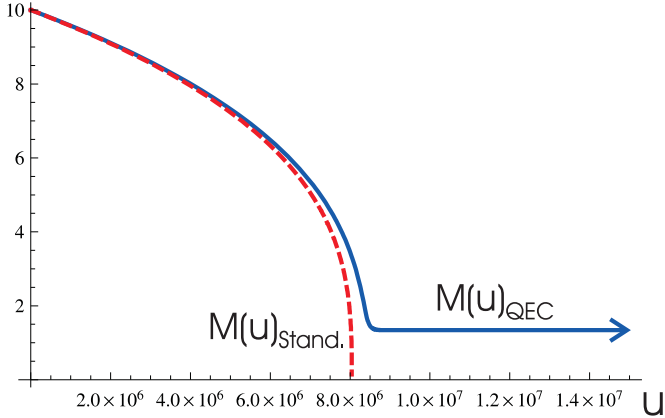}
\caption{\label{Massevol} In this figure we compare the evolution of a black hole mass without taking into account quantum corrections and energy conservation $M(u)_{Stand.}$ (dashed line) and taking them into account $M(u)_{QEC}$  (solid line) in both cases starting with the same initial mass at $u=0$. The formation of a remnant, if quantum corrections and energy conservation are taken into account, contrasts with the sudden total evaporation in the standard `thermal' approximation.}
\end{figure}

On the other hand, if we want to take into account quantum corrections and energy conservation (including its corresponding modified backscattering) it will be necessary to use expression (\ref{lumi}) for the luminosity.
From (\ref{lumi}) and (\ref{difM}), one finds that in order to find the evolution of the black hole mass one has to solve the differential equation
\begin{equation}
\frac{d M(u)}{du}=- \frac{1}{2 \pi}\int_0^{M(u)} \frac{4 E^3 R_+(E)^2}{\exp\left(4\pi \int_{0}^{E}  \frac{dE'}{g(R_+;E')}\right)-1} dE.
\end{equation}
The results of the numerical integration are shown in figure \ref{Massevol}, where the evolution of the mass of the quantum black hole taking into account backscattering and energy conservation is also compared with the previous `thermal' result (\ref{MassScat}). In general, starting from the same initial macroscopic mass (above the critical mass), the evaporation process is slower than in the standard thermal case if quantum corrections and energy conservation are taken into account.
This is due to the lower luminosity in the quantum corrected case or, in other words, to its smaller temperature (see (\ref{TempAprox}) for $M\gg m_p$). In the lasts stages of evaporation, as the critical mass is reached, the luminosity tends to zero if quantum modifications and energy conservation are taken into account. Specifically, using (\ref{lumi}) the luminosity around the critical mass (see figure \ref{luminos}) can be approximately described by
\begin{equation}\label{Laprox}
L(M)= k(\gamma) (M-M_{cr})^{7/2},
\end{equation}
where, for instance, we have for $\gamma=0$
\[
k(0)=\frac{17 \sqrt{2} G_0^{3/4} \tilde{\omega}^{1/4}}{140 \pi^2 }.
\]
If one now uses (\ref{Laprox}) into the differential equation (\ref{difM}) with the initial condition $M(u=0)=M_0$, one gets for the evolution of the mass around its critical value the decay
\begin{equation}
M(u)=M_{cr}+\frac{M_0-M_{cr}}{\left[1+\frac{5}{2} k (M_0-M_{cr})^{5/2} u\right]^{2/5}},
\end{equation}
satisfying $M(u\rightarrow \infty)=M_{cr}$. In other words, a remnant is formed with a mass of the order of Planck's mass ($M_{cr}$). The reader should note that the result of the formation of a remnant does not depend on the necessary consideration of the backscattered radiation since, in case of disregarding it ($\beta_{EC}=1$), the luminosity (\ref{lumi}) around the critical mass would have been approximately $L(M)\propto (M-M_{cr})^{3/2}$. So that, using (\ref{difM}), we would have again obtained the formation of a remnant. Therefore, just by including quantum correction to the effective metric one obtains a picture that  necessarily contrasts with the standard thermal picture in which the evaporation accelerates until reaching a sudden total evaporation.

\section{Correlations among the emitted particles}\label{secescape}

In this section we will try to verify that, taking into account energy conservation in the tunneling process, the emitted particles
are correlated, allowing thus the information to escape from the black hole as it evaporates. We can expect this result from
the non-thermal behavior found above (\ref{emprob}). Following the analysis showed in \cite{Z1,Z2,I&Y}
where the authors linked the existence of correlations among tunneled particles and the entropy conservation of the full
system (black hole plus Hawking radiation),
we have to compute the statistical correlation $C(E_1,E_2)$ between an event consisting in the emission of a quantum of energy $E_1+E_2$ and an event consisting in the emission of a quantum of energy $E_1$ \emph{and} a quantum of energy $E_2$.
Using equation (\ref{emprob})
for the emission probability
\begin{eqnarray}
 \label{correlations}
C(E_1,E_2) &= & \ln\mid\Gamma(E_{1}+E_{2})\mid-\ln\mid\Gamma(E_{1})\Gamma(E_{2})\mid = \nonumber \\
&= & \Delta\mathcal{S}(E_1+E_2)-\Delta\mathcal{S}(E_1)-\Delta\mathcal{S}(E_2),
\end{eqnarray}
where
\begin{eqnarray}
\Delta\mathcal{S}(E)\equiv-2\, \mbox{Im S}=-4\pi \int_{0}^{E}  \frac{dE'}{g(R_+;E')}
\end{eqnarray}
is the change in the hole's entropy \cite{P&W}\cite{I&Y}.
Note that the fact that (\ref{correlations}) does not have to be zero implies that the imposition of energy conservation lead us to the existence of correlations among tunneled particles and, in this way, the information could escape from the black hole.

In fact, we can distinguish two different regimes. In the one hand, we can consider the regime in which the black hole mass is much bigger than Planck's mass. Then we have
\begin{equation}
C(E_1,E_2) =  8\pi G_{0}E_{1}E_{2}
-2\pi(1+\gamma)\;\tilde{\omega}\;\ln\left[\frac{(M-E_{1})(M-E_{2})}{M(M-E_{1}-E_{2})}\right]+{\cal O}(\tilde{\omega}^2) \,.
\end{equation}
We see that if we turn off the quantum corrections ($\tilde{\omega}=0$) one recovers the standard result for the correlation \cite{Z1,Z2,I&Y}. However, the quantum correction provide us with a slightly smaller statistical correlation.
On the other hand, if the black hole mass is of the order of the critical mass (and, thus, of Planck's mass) we enter a regime in which quantum corrections become essential. In particular, if we take into account that $E\leq M-M_{cr}$, we have (as was seen for the calculations of the luminosity)
\begin{displaymath}
\lim_{M\rightarrow M_{cr}} \int_0^E \frac{dE'}{g(R_+;E')}=0
\end{displaymath}
and, taking into account (\ref{correlations}),
\begin{displaymath}
\lim_{M\rightarrow M_{cr}} C(E_1,E_2)=0.
\end{displaymath}
Summarizing, thanks to energy conservation information can escape a macroscopic black hole during its evaporation. However as its mass approaches the critical mass only long wavelength massless particles can tunnel out of its horizon carrying an always decreasing amount of information.

\section{Summary and conclusions}\label{seccon}

In this letter we have considered the radiation coming out from a quantum black hole when energy conservation is satisfied. This has been accomplished by using the tunneling method proposed in \cite{P&W} (corresponding with the usual picture of pair creation near the horizon and the subsequent tunneling of particles) to a quantum black hole derived in the Quantum Einstein Gravity framework. It is important to remark that, despite the study has been carried out for generic `improved' quantum black holes, we expect that both the effective black hole solution and the tunneling method should be of limited reliability for black holes around the Planck regime.

First, we have seen that the tunneling method leads to a deviation from an exact thermal spectrum when it is applied to the improved black hole, what is reflected in the expression (\ref{emprob}). However, only in the approximation of emission of low-energy particles (i.e., neglecting second order energy terms in (\ref{emprob})), we have been able to assign a quantum corrected temperature for the black hole. This temperature is always smaller than the Hawking-Bekenstein temperature and it even becomes zero for a black hole of mass equal to the critical mass.

Contrary to the thermal emission, we have seen that the tunneling method provide us with a limit for the energy carried out by a particle emitted from the black hole. In particular, we have found the satisfactory result that, in agreement with energy conservation, the emitted quantum must have an energy smaller than the black hole's mass. More specifically, the energy $E$ of the quantum should satisfy $E\leq M-M_{cr}$.

In order to study the evolution of the improved black hole as it evaporates respecting energy conservation, we have accordingly also modified the grey-body factor, which allows to consider the effect of the emitted radiation that is later backscattered. We would like to emphasize that the effect of the backscattering can not be overlooked since this would result in luminosities around ten times bigger for macroscopic black holes and the subsequent much faster evolution.

Equipped with the non-thermal distribution and the modified grey-body factor we have been able to derive an expression for the luminosity (\ref{lumi}) that takes into account quantum corrections as well as the full consequences of energy conservation. The standard `thermal' result implies a diverging luminosity as the black hole's mass tends to zero. A result that comes directly from the violation of energy conservation. However, by taking into account quantum corrections and energy conservation we have shown that the luminosity reaches a maximum for a non null value of the black hole mass. It even would be zero for a black hole with a mass equal to the critical mass.

We have studied the evaporation process taking into account the back-reaction on the metric caused by the decrease in the black hole mass. The evolution has been compared with the result in the standard `thermal' approximation showing that
the process is slower when quantum corrections are taken into account. In particular,
the mass decays in the lasts stages of the evaporation ever approaching the critical mass and, thus, creating a remnant.

Finally, since the claim of \textit{information loss} in black holes has as one of its pillars that black holes have an exact thermal spectrum, we have analyzed the consequences that the deviation from thermality in the quantum corrected black hole could have for the escape of information. Following the careful method put forward in \cite{Z1,Z2,I&Y} we have studied the correlations among the emitted particles showing that, contrary to the thermal case, the correlations exist allowing the information to escape. However, in the lasts stages of the evaporation, as the black hole's mass approaches the critical mass, only long wavelength particles can tunnel out the horizon limiting drastically the amount of information that can be carried away from the black hole.



\end{document}